\documentclass[12pt]{article}
\begin{document}
\baselineskip 18pt
\newcommand{\Tr}{\mbox{Tr\,}}
\newcommand{\beq}{\begin{equation}}
\newcommand{\eeq}[1]{\label{#1}\end{equation}}
\newcommand{\bea}{\begin{eqnarray}}
\newcommand{\eea}[1]{\label{#1}\end{eqnarray}}
\renewcommand{\Re}{\mbox{Re}\,}
\renewcommand{\Im}{\mbox{Im}\,}
\begin{titlepage}
\hfill Bicocca-FT/99/27\vskip .01in \hfill 
NYU-TH/99/09/02\vskip .01in \hfill hep-th/9909047
\begin{center}
\hfill
\vskip .4in
{\large\bf The Supergravity Dual of $N=1$ Super Yang-Mills Theory}
\end{center}
\vskip .4in
\begin{center}
{\large L. Girardello$^{a}$, M. Petrini$^b$, M. Porrati$^{c}$ and
A. Zaffaroni$^a$\footnotemark}
\footnotetext{e-mail: luciano.girardello@mi.infn.it, m.petrini@ic.ac.uk,
massimo.porrati@nyu.edu, alberto.zaffa-\\ roni@cern.ch}
\vskip .1in
(a){\em Dipartimento di Fisica, Universit{\`a} di Milano-Bicocca and INFN,
Sezione di Milano, Italy}
\vskip .1in
(b){\em Theoretical Physics Group, Blackett Laboratory,
Imperial College, London SW7 2BZ, U.K.}
\vskip .1in
(c){\em Department of Physics, NYU, 4 Washington Pl.,
New York, NY 10003, USA}
\end{center}
\vskip .4in
\begin{center} {\bf ABSTRACT} \end{center}
\begin{quotation}
\noindent
We find an exact, $N=1$ supersymmetric kink solution of 5d gauged
supergravity. We associate this solution with the RG flow from 
$N=4$ super Yang-Mills theory, deformed by a relevant operator, to
pure $N=1$ super Yang-Mills in the IR. We test this identification by
computing various QFT quantities using the supergravity dual:  the
tension of 
electric and magnetic strings and the gaugino condensate. 
As demanded by our identification, our kink solution is a
true deformation of $N=4$, that exhibits
confinement of quarks, magnetic screening, and spontaneous chiral symmetry
breaking. 
\end{quotation}
\vfill
\end{titlepage}
\eject
\noindent
\section{Introduction}
In previous papers~\cite{gppz1,gppz2} we explored several properties of
relevant deformations of strongly-coupled $N=4$ super Yang-Mills using
its dual~\cite{malda,gkp,w1} description in terms of
5d gauged supergravity.
In this paper we present a holographic RG flow from $N=4$ SYM to pure
$N=1$
SYM in the IR. We find agreement with field theory
expectations, namely, quarks confine, monopoles are screened, and there is
a gaugino condensate.

This paper is the natural continuation of \cite{gppz1,gppz2}, where we
studied the supergravity description of RG flows.
In particular, in~\cite{gppz1} we proved the existence of a kink solution
of 5d supergravity that was interpreted as the dual of a mass-deformed
$N=4$ super Yang-Mills, flowing in the IR to a
non-supersymmetric, interacting, local conformal field theory (CFT)
(see also~\cite{dz}).
We also introduced a c-function and we proved a c-theorem valid for all
supergravity flows involving the metric and an arbitrary number of
scalars. Such a setup, namely coupled gravity-scalar equations is remarkably
versatile, and it can also describe deformations of $N=4$ flowing to
non-conformal, interacting theories.

Most of the solutions asymptotic to the $N=4$ UV fixed point
go to infinity in the space of parameters described by the 5d supergravity.
It is natural to interpret
these solutions as RG flows to non-conformal IR theories.
Examples were considered
in \cite{gppz2,freed2}. The generic solutions discussed in \cite{gppz2}
exhibit different behaviours in the IR, from confinement to screening,
associated with an IR singularity that is common to many 10d solutions
\cite{ks,gubser,myers,pol,yank,Od}. 
In \cite{freed2} the Coulomb branch of
$N=4$ YM was studied.

In view of the double role
of supergravity solutions with an asymptotic AdS region (deformations of
an UV fixed point versus 
the same theory in a different vacuum \cite{bala,kw2}),
we may ask
whether the solutions in \cite{gppz2} are true deformations of $N=4$ YM
or simply different states (e.g. spontaneously broken vacua)
of the same theory.
The case of \cite{ks,gubser}
turned out to correspond to a different vacuum of $N=4$ YM; the same
is certainly true also for
\cite{freed2}. Unlike the solutions in~\cite{freed2,ks,gubser}, the ones
in~\cite{gppz2}  are expected to describe
deformations. However,
up to now, no explicit, analytically solvable example 
of deformations was given in literature.
One of the purposes
of this paper is to provide such an example, describing the pure $N=1$ YM case.

Another important question is the reliability of supergravity solutions.
Problems were pointed out in \cite{freed2},
after an explicit comparison
of supergravity predictions to gauge theory results.
In the screening case discussed in \cite{freed2},
the type IIB string goes near the singularity,
and any actual computation requires the form of the solution in the IR,
where large corrections to the supergravity result are expected.
It is then important, to better understand the seriousness of this problem,
to consider other solutions where comparison with quantum field theory
expectation can be made. The simplest such case,
which corresponds to a deformation of the $N=4$ theory, is the flow
to pure $N=1$ YM.
In this paper, 
we exhibit a solution with all qualitative features of pure $N=1$ super YM 
theory, in particular, quark confinement.
 Even in our case the confining string
probes a high-curvature singularity; it is therefore encouraging to find that
at least in some cases this problem does not change the qualitative picture
(confinement, magnetic screening, chiral symmetry braking) emerging
from field theory.
To better control our solutions we are led to consider supersymmetric
examples. The equally interesting $N=0$ solutions are certainly less
reliable, even though they are easier to study \cite{gppz2}.

This paper is organised as follows:
in Section 2, we give a brief review of the basic results
in the interpretation of 5d supergravity solutions as
RG flows, following \cite{gppz1,gppz2}.
We also discuss the cases of flows to IR conformal and non-conformal
theories, with particular attention to open problems. Section 3 describes
in details the flow to $N=1$ YM and its properties. The computation of
electric and magnetic strings, and the identification of
the supergravity field describing the gaugino condensate are 
described. That Section also contains a brief discussion of the stability
of chirally-symmetric versus chiral-symmetry breaking vacua.
The Appendix contains various useful formulae used throughout the paper.

\section{RG Flow from 5d Supergravity: a Brief Review}
To fix once for all our notations, we will use coordinates where
the Anti-de-Sitter space reads
\beq
ds^2= dy^2 + e^{2y/R}dx^\mu d x_\mu , \;\;\;\mu =0,1,2,3.
\eeq{I1}
$R$ is the $AdS$ radius and it will be henceforth set to $1$, unless
explicitly stated.

The fifth coordinate $y$ of $AdS_5$ has a natural interpretation as an energy
scale \cite{malda,PP,ps,bk}. As in~\cite{gppz1}, we associate larger
energies with increasing $y$.
We look for the supergravity description of the quantum field theory
RG flow that connects, say, the $N=4$ YM theory to some IR CFT.
It was proposed in \cite{gppz1} to identify this RG flow
with solutions of type IIB supergravity
that interpolate between $AdS_5\times S^5$ at $y=\infty$ and the
 type IIB background associated with the IR CFT at $y=-\infty$.
We are interested in studying relevant deformations of the $N=4$ theory.
The explicit dependence on $y$ must therefore break the conformal group
$O(4,2)$ to the 4d Poincar{\'e} group.
$y=\infty$ will be considered the UV, while $y=-\infty$ the extreme IR.

In the previous and following discussions, the $N=4$ theory can be
replaced by any other CFT that admits an $AdS$ description.

Supergravity solutions have a double meaning: they may describe
deformations of a CFT, or, the same theory in a different vacuum
\cite{bala,kw2}.
Indeed, the running of coupling constants and parameters along the RG flow
can be induced in the UV theory in two different ways:
by deforming the CFT with a
relevant operator, or by giving a nonzero VEV to some operator. 
What the case is for a given solution
depends on the asymptotic UV behaviour.
The CFT operators
that are used as deformations, or that are acquiring a non-zero VEV, can be
found by linearising the solution near $y=\infty$. The
type IIB supergravity scalar modes $\lambda(y)$ that deform
the $AdS_5\times S^5$ solution at $y=\infty$ are associated with
CFT operators $O_{\lambda}$ \cite{gkp,w1}. The linearised 5d equation of motion
for a fluctuation $\lambda(y)$ in the $AdS$-background eq.~(\ref{I1}) reads
\beq
\ddot\lambda +4\dot\lambda=M^2\lambda,
\eeq{I2}
where the dot means the derivative with respect to $y$.
The mass of the supergravity mode is related to
the conformal dimension $\Delta$ of the operator $O_\lambda$ by
$\Delta=-2 +\sqrt{4 + M^2}$ \cite{gkp,w1}.

Equation~(\ref{I2}) has a solution depending on two arbitrary
parameters
\beq
\lambda(y)=Ae^{-(4-\Delta )y}+Be^{-\Delta y}.
\eeq{I3}
We are interested in the case of relevant operators, where $\Delta\le 4$.
A solution with $A\ne 0$ is (UV) asymptotic to the
non-normalizable solution $e^{-(4-\Delta )y}$
\cite{w1}, and is associated with a deformation of the $N=4$ theory
with the operator $O_\lambda$. On the other hand, a solution asymptotic
to the normalizable\footnote{Here we are only considering the UV behaviour.}
solution $e^{-\Delta y}$ ($A=0$) is associated with a different
vacuum of the UV theory, where the operator $O_\lambda$ has a non-zero VEV
\cite{bala,kw2}.

\subsection{The 5d Supergravity: Flow Between CFT's and the c-Theorem}
10d solutions interpolating between CFT's are difficult to find because of the
high non-linearity of the type IIB equations of motion.
Perturbation theory around the UV point is not practicable and
exact solutions are not easy to construct.
  If we restrict
the number of operators that we may use to deform the UV
fixed point, we can consider the 5d gauged supergravity instead of
the full 10d theory \cite{gppz1,dz}.

The $N=8$ gauged supergravity \cite{5dgauged}
is a consistent
truncation of type IIB on $S^5$ \cite{trunc}. This
means that every solution of the 5d theory can be lifted to a consistent 10d
type IIB solution.
The operators that can be described using 5d supergravity are enough
to mimic many RG flows. In particular, all mass terms of fermions and scalars,
which can be described by supergravity modes, are included in the 5d
truncated theory. The only relevant operators that cannot be described
by this method are cubic terms in the scalars. Five-dimensional
gauged supergravity
has 42 scalars, which
transform under the $N=4$ YM R-symmetry $SU(4)$ as
$\underline{1},\underline{20},\underline{10}$.
The singlet is associated with the marginal deformation
corresponding to a change in the coupling constant of the $N=4$ theory.
The $\underline{20}$ is a mode with mass square $M^2=-4$ and is associated with
the symmetric traceless operator $\Tr\phi_i\phi_j, i,j=1,...,6$ of dimension 2.
The $\underline{10}$ has mass square $M^2=-3$ and is associated with
the fermion bi-linear operator $\Tr\lambda_A\lambda_B, A,B=1,...,4$, of
dimension
3. Using these operators, we can discuss at least all mass deformations
that have a supergravity description\footnote{The only missing state,
the R-symmetry singlet $\sum_{i=1}^6\Tr\phi_i\phi_i$, is associated with
a stringy mode.}.

The effective 5d Lagrangian, restricted to the 42 scalars $\lambda_a$, is
\beq
L= \sqrt{-g}\left [ -{R\over 4}+ {1\over 2}g^{IJ}\partial_I\lambda_a \partial_J
\lambda_b G^{ab} + V(\lambda)\right ],
\eeq{I4}
In all our applications, we will choose a convenient parametrisation
where the scalar kinetic term is canonical ($G^{ab}=\delta^{ab}$).
This is the most general Lagrangian for scalars coupled to gravity. The
contribution of the 5d gauged supergravity \cite{5dgauged} is to provide
the potential $V(\lambda)$.

$V(\lambda)$ has a central critical point where all the $\lambda_a$ vanish
with $SO(6)$ symmetry and $AdS$ metric: it corresponds to the $N=4$ YM theory.
It is natural to associate all other $AdS$ critical points of the potential
with IR CFT's that can be connected to the $N=4$ theory by an RG flow.
The scalars with a non-zero VEV, $\lambda_{IR}$,
at the critical points indicate which
operators have been used to deform the UV theory. Up to now,
all critical points with at least $SU(2)$ symmetry have been classified
\cite{grw2,pilch}.
There exist three $N=0$ theories with symmetry $SU(3)\times U(1)$, $SO(5)$ and
$SU(2)\times U(1)^2$. All those IR CFT theories are obtained as mass
deformations of the $N=4$ theory \cite{gppz1,dz}.
Both the first two critical points might be
unstable~\cite{dz,dzf}; the third one has not been yet investigated. Moreover,
there
is an $N=1$ point with symmetry $SU(2)\times U(1)$ \cite{pilch},
which is stable due to
supersymmetry; it is associated with the $N=4$ theory deformed with
a supersymmetric mass for one of the three $N=1$ chiral superfields
\cite{lust}. The value of the potential at the critical point (determining the
cosmological constant of the $AdS$ space) is associated with the central
charges $c=a$ of the CFT by~\cite{henningson}
\beq
c=a\sim (-V(\lambda_{IR}))^{-3/2}.
\eeq{I5}
In the case of the supersymmetric point $SU(2)\times U(1)$ the central charge
predicted by supergravity agrees with quantum field theory
expectation \cite{lust}\footnote{Supersymmetry allows to compute $c$ and $a$
using the results in \cite{anselmi}.}.

The supergravity description of the RG flow connecting the $N=4$ theory
to one of these IR CFT's was found in \cite{gppz1} in the form of a kink
solution, which interpolates between the two critical points. The form of
a 4d Poincar{\'e} invariant metric is
\beq
ds^2= dy^2 + e^{2\phi(y)}dx^\mu d x_\mu , \;\;\;\mu =0,1,2,3.
\eeq{I6}
We look for solutions that are asymptotic to $AdS_5$ spaces (but with
different radii $R_{UV}$ and $R_{IR}$) both for $y\rightarrow\infty$ and
$y\rightarrow-\infty$: $\phi(y)\rightarrow y/R_{UV}+\hbox{const}$ for
$y\rightarrow\infty$, and  $\phi(y)\rightarrow y/R_{IR}+\hbox{const}$ for
$y\rightarrow -\infty$. Analogously, $\lambda(y)\rightarrow 0$ for
$y\rightarrow\infty$, while $\lambda(y)\rightarrow \lambda_{IR}$ for
$y\rightarrow -\infty$. The equations of motion 
for the scalars and the metric read
\bea
 \ddot\lambda_a+4\dot\phi\dot\lambda={\partial V\over\partial\lambda_a},\\
6(\dot\phi )^2=\sum_a(\dot\lambda_a )^2 -2V.
\eea{I7}
With the above boundary conditions
and a reasonable shape for the potential,
a kink interpolating between critical points always exists \cite{gppz1}.
The argument is roughly as follows: by eliminating $\phi$ from one
of the equations, the problem reduces to the classical motion
of a particle in a potential with never-vanishing, positive damping.
Since the particle is rolling down from the top of a hill (the IR CFT),
the damping implies that it will stop at the bottom of the hill
(the $N=4$ point)\footnote{Note that, in this particle interpretation,
 we are going from IR to UV.}.

The interpretation of the kink as a RG flow can be strengthened 
by exhibiting a candidate c-function \cite{gppz1},
\beq
c(y)= {\rm const\,} (T_{yy})^{-3/2}.
\eeq{I8}
constructed with the $y$ component of the stress-energy tensor,
$T_{yy}=6(\dot\phi )^2=\sum_a(\dot\lambda_a )^2 -2V$. This expression
certainly reduces to the central charge eq.~(\ref{I5}) at the critical points,
where $\dot\lambda_a=0$. Moreover, it is a straightforward exercise
to prove that $c(y)$ is monotonic, using the equations of motion and the
already-mentioned 
boundary conditions\footnote{From the equations of motion we have
$6\ddot\phi=-4\sum_a(\dot\lambda_a )^2$ and it is easy to check that,
due to the boundary conditions, $\dot\phi$ is always positive. To
avoid confusion, note
that $y$ measures the energy in a direction opposite to the one of
the RG group: $y$ increases when the energy increases. $c(y)$ is however
monotonic exactly in the expected way: it decreases going from UV to IR.}.
Even more simply, $c(y)$ is 
inversely proportional to the energy of our classical
particle in a potential, and, with a positive damping, energy always
decreases. This proof of the c-theorem is valid for all the
supergravity
flows involving the metric and an arbitrary number of scalars; it 
generalises to generic Poincar{\'e} invariant supergravity
solutions in \cite{freed1}.

In ref.~\cite{freed1} the conditions for a supersymmetric flow were found.
As usual, a solution for which the fermionic shifts
vanish, automatically satisfies the equations of motion. Moreover, this
shortcut reduces the second order equations to first order ones.
For a supersymmetric solution, the potential $V$
can be written in terms of a superpotential, $W$, as
\beq
V=\frac{g^2}{8}\sum_{a=1}^{n}\left|
\frac{\partial W}{\partial \lambda_a} \right|^2
- \frac{g^2}{3} \left|W \right|^2,
\eeq{I9}
where $W$ is one of the eigenvalue of the tensor $W_{ab}$ defined in
\cite{grw2}. The equations of motion reduce to
\bea
\dot\lambda_a&=&\frac{g}{2} \frac{\partial W}{\partial \lambda_a},\\
\dot\phi&=& - \frac{g}{3} W.
\eea{I10}
It is easy to check that a solution of eq.(\ref{I10}) satisfies
 also the second order equations ~(\ref{I7}). Using this result,
in \cite{freed1} the $N=1$ fixed point $SU(2)\times U(1)$ was studied
in detail, finding agreement with quantum field theory expectations.

All these kink solutions generically correspond to deformations of
the UV fixed point; this can be explicitly checked for the $SU(2)\times U(1)$
point.


\subsection{Flow to Non-Conformal Theories}
 What remains yet
to discuss is the most interesting part of the story, but
also the subtlest:
the case of  non-conformal IR theories.
Most solutions that start at the $N=4$ fixed point do not
flow to a different critical point, instead, they run away to infinity in
configuration space.
It is natural to interpret these solutions as RG flows to non-conformal
theories \cite{gppz2}. It is indeed expected that most of the mass
deformations of $N=4$ fixed point flow to pure-glue theories in the IR.
In \cite{gppz2}, we examined a general class of solutions
and we provided the general
tools for studying the IR behaviour of those theories. Solutions were studied
for which the scalar potential becomes irrelevant in the IR. Despite
this restriction, those solutions can be called ``generic,'' since they
fill a large subset of the parameter space.
In those solutions the IR 5d Einstein metric has a ``universal'' singular form
\beq
ds^2= dy^2 + \sqrt{y-a}dx^\mu d x_\mu , \;\;\;\mu =0,1,2,3,
\eeq{I11}
with a logarithmic behaviour $\lambda_a=A_a \log (y-a)$ for the scalar fields.

Supersymmetric solutions are not ``generic,'' since they satisfy
eq.~(\ref{I10}), which prevents the potential from being irrelevant
in the IR. However, modifications of the results
of \cite{gppz2} due to supersymmetry are minimal, mostly a slight change in
the form of the IR metric, which remains singular; scalar fields
maintain a logarithmic behaviour.
Supersymmetric solutions associated with the Coulomb branch of
the $N=4$ theory were studied in \cite{freed2}.

In the solutions found in \cite{gppz2}, there is a singularity in the IR,
whose generic form
is common to many ten dimensional
solutions from type IIB \cite{ks,gubser, myers} to type OB \cite{minahan} and
even to
non-critical strings \cite{pol,yank}.
Agreement with the solutions in \cite{ks,gubser,myers} is expected, since they
can be described in the language of $N=8$ 5d gauge supergravity. Agreement
with type O and non-critical strings is unexpected, but it becomes a little
less surprising when one realizes that those solutions can be reduced
to effective 5d theories of scalars coupled to gravity.

Depending on the value of the parameters $A_a$, the solutions may exhibit
conformal behaviour, confinement, or screening in the
IR. The appropriate case can be determined by computing a Wilson loop.
Following \cite{maldaloop}, the Wilson
loop associated with a contour $C$ on the boundary is
obtained minimising the world-sheet action of a string living in the
background (\ref{I6}) and whose endpoints
live on the boundary contour $C$.
By considering a rectangular loop $C$, with one side of length $L$
in the direction $x$ and another one of length $T$ along the time
axis, and by choosing the standard embedding $\sigma
=x, \tau=t$ we find:
\beq
S= \int d\tau d\sigma \sqrt{G_{ind}} = T\int dx
T(y)e^{\phi(y)}\sqrt{(\partial_x y)^2 + e^{2\phi(y)}}.
\eeq{I12}

Here $T(y)$ is the tension of the fundamental (the case of a quark)
or of the D1 string (monopole) in five
dimensions, which in general is a non-trivial function of
the scalar fields.
As discussed in \cite{gppz2} on the basis of
 naive dimensional reduction from ten dimensions, the tensions can be
read from the coefficients of the kinetic term for the NS-NS and R-R
antisymmetric tensors, respectively. Since in the
five-dimensional supergravity Lagrangian \cite{grw2} the kinetic term
for the antisymmetric tensors is written in the first order formalism
\beq
\epsilon_{\alpha\beta}B_{I\alpha}\wedge dB_{I\beta} + A_{I\alpha,J\beta}
B_{I\alpha}\wedge *B_{J\beta},
\eeq{I13}

the tensions come from the matrix $A_{I\alpha,J\beta}$ in the quadratic
term. In particular they receive contribution only from the diagonal
entries of $A$, the off-diagonal ones contributing to the
mass terms for the B tensors.

The Wilson loop is conveniently studied in the coordinates where
the quark-antiquark (or monopole-antimonopole) energy reads,
\beq
E = S/T = \int dx \sqrt{(\partial_x u)^2+f(u)},
\eeq{I14}
the change of variable being given by
\beq
{\partial u\over \partial y}=T(y)e^{\phi(y)},\,\,\, f(u) = T^2(u)e^{4\phi(u)}.
\eeq{I15}
The boundary is now at $u=+
\infty$, while the IR region corresponds to small values of $u$.

The Wilson loop behaviour is determined by that
of the function $f(u)$ \cite{gppz2,brand}. \cite{gppz2} reviews in detail
all the
different cases that can occur, here we give only a sketch of the results.
If $f(u)\sim u^\gamma$   for small $u$, with $\gamma\ge 2$
the quark-antiquark energy
has a power-like behaviour, with exponent depending on $\gamma$.
The conformal case is $\gamma=4$. If $0\le \gamma\le 2$, then the string
will find energetically favourable to go straight to the IR region,
where to separate the quarks does not cost any energy \cite{brand}.
When $\gamma\le 0$, there is an effective barrier that prevents
the string
from penetrating into the deep IR
region; it will live near the minimum of the function $f(u)$.
We therefore expect an area law behaviour, i.e. confinement.
A similar result holds when the function $f(u)$ does not diverge, but
is strictly bounded above zero. The case of a barrier (confinement)
is the one where the supergravity computation is more reliable.
It is plausible that the  minimum of $f(u)$ lives in a region far
from the singularity, where supergravity can be still trusted.
On the other hand, the case of screening is the least reliable. The string
has to touch the singularity before the quarks can be separated
without spending energy; but large corrections to the supergravity result
certainly are expected
in the deep IR region.

Having settled all the machinery for studying solutions where scalars run 
to infinity,
we may start to discuss their physical meaning. The first important problem
is to understand whether a solution corresponds to a deformation or
 to a different vacuum of the $N=4$ theory.
After initial confusion, the solution in \cite{ks,gubser} was identified
as a different phase of the $N=4$ theory; the same is certainly true
for the solutions in \cite{freed2}, describing the Coulomb branch of $N=4$ YM.
We should not conclude from this that all the solutions in \cite{gppz2}
correspond to vacua of the $N=4$ theory.

The solutions in \cite{freed2,ks,gubser,myers,tseytlin} are associated with
operators that may reasonably have a non-zero VEV in the $N=4$ theory
\footnote{We can understand which operator gets a VEV or appears as a
deformation
by looking at
the supergravity fields that run in $y$.}.
It is  reasonable to
associate solutions were the dilaton (and/or axion)
runs \cite{ks,gubser, tseytlin}
with vacua with non-zero $\langle\Tr F^2\rangle$, and solutions where the
scalars in the
$\underline{20}$ run \cite{freed2} with the Coulomb branch of $N=4$ YM,
where $\langle\Tr(\phi_i\phi_j)\rangle\ne 0$. It is instead difficult to image
solutions
where the $\underline{10}$ runs as vacua of the $N=4$ theory
with non-zero fermionic condensate. It is quite plausible that many
of the solutions with non-zero $\underline{10}$ correspond to deformations of
the UV fixed point rather than different vacua. Since, in general,
 we do not know the full
solution connecting the UV with the singular IR region, we can not
explicitly check this claim.
Here, we simply note that the description of a different vacuum
requires a UV-normalizable solution, with asymptotic behaviour as in
eq.~(\ref{I3}).
This would require a fine tuning ($A=0$); obviously, this
cannot be the generic situation.
Note also
that the case of a running
dilaton considered in \cite{ks,gubser}, which turned out to correspond
to a vacuum of the theory, is quite special. For operators of dimension 4,
equation~(\ref{I3}) becomes,
\beq
\lambda(y)=A+Be^{-4y}
\eeq{I16}
The non-normalizable solution, the one associated with
a true deformation of the UV theory, here becomes a constant. In the
case of the dilaton, a non-normalizable solution can be
always re-absorbed into a shift of the $N=4$ coupling constant.
This means that every solution where only the dilaton (and/or axion)
runs
is UV-asymptotic to the normalizable solution; therefore, it
corresponds to a different vacuum of the $N=4$ theory. Completely different
is the case of dimension-three operators, where both behaviours
in eq.~(\ref{I3}) are allowed.
In this paper we provide an explicit example of solution associated with
a deformation of the $N=4$ theory.

The UV behaviour in the supersymmetric case can always be determined
using equations ~(\ref{I10}).
In this paper we will consider a supersymmetric case, the RG flow from
$N=4$  to pure $N=1$ YM, which corresponds to the running of fields
in the $\underline{10}$; it can be explicitly recognised
as a deformation of the UV theory. We will prove that quarks confine and
monopoles are screened by computing a Wilson loop.
Once we are in the the $N=1$ theory, the existence of a fermionic condensate
is no more to be regarded as suspect; it is instead expected, since
pure $N=1$ YM has a gluino condensate. We will check the existence
of the condensate by playing with the double role of supergravity solutions,
deformations versus VEV. We will find a supersymmetric solution of
the equations of motion depending on two scalars $m$ and $\sigma$,
the first being associated with the mass term which breaks $N=4$ to $N=1$
pure YM, the second being associated with the gluino condensate operator
\footnote{These fields appear as $\underline{6}$ and $\underline{1}$ in the
decomposition $\underline{10}\rightarrow
\underline{1}+\underline{6} +\underline{3}$ of $SU(4)$
under $SU(3)\times U(1)$.}. By checking the asymptotic UV behaviour,
$m$ turns out to be associated with a deformation of the $N=4$ theory,
while $\sigma$ with a VEV. This means that our solution describes
an RG flow to the $N=1$ pure YM in a vacuum with non-zero condensate.

We must spend a few words on the reliability of these singular solutions.  
Since the curvature and the kinetic terms for the scalars
typically diverge in the IR region, large corrections to supergravity
may be expected. An explicit comparison between supergravity and
quantum field theory has been made in \cite{freed2} for the case of the $N=4$
Coulomb branch and several discrepancies were pointed out.
By itself, this result does not immediately
implies that all the other solutions
describing confinement should be considered as suspect.
In a certain sense, screening is more sensitive to
infrared physics than confinement: any amount of string tension, no matter how
small, will destroy it.

Our contribution to the debate about reliability of supergravity solutions is
to present a pure $N=1$ YM example, where all qualitative
features expected in the quantum field theory, namely confinement and
a gaugino condensate, can be explicitly found in the supergravity side.
We must also point out that the singularity of the metric eq.~(\ref{I11})
might be an artifact of the 5d Einstein-frame metric. Such a singularity may
appear also when the higher-dimensional metric is regular. For instance,
write $R^9$ with flat metric in polar coordinates as
$R^+\times S^3\times R^5$, perform a dimensional reduction to
$R^+\times R^5$, and rescale to the 5d Einstein frame: the resulting metric
has the same singularity (at $r=0$) as in eq.~(\ref{I11}).
Naturally, when the singularity is an artifact of the dimensional reduction,
the supergravity computation of the string tension is completely justified.
This is the case, for instance, of the non-conformal backgrounds in
ref.~\cite{w3}
\section{The $N=1$ Theory}



In this Section we will discuss the properties of the five-dimensional
supergravity solution corresponding to
a deformation of $N=4$ Super Yang-Mills
theory with a supersymmetric mass term for the three
fermions in the chiral $N=1$ multiplets. In $N=1$ notation, this is a
mass
term for the three chiral superfields $X_i$
\beq
\int d^2\theta m_{ij} \Tr X_i X_j + {\rm c.c.},
\eeq{m31}
where $m_{ij}$ is a complex, symmetric matrix.

The theory flows in the IR to pure $N=1$ Yang-Mills, which confines.

To obtain the standard $N=1$ pure Yang-Mills with fixed scale
$\Lambda$, we need a fine
tuning of the UV parameters, in which the mass $m$ diverges while
the 't Hooft
coupling constant, $x$, goes to zero as an (inverse) logarithm of $m$.
This is outside
the regime of validity of supergravity, which requires a large $x$,
but we may still expect to find from supergravity the qualitative
properties of the theory.

\subsection{Mass Deformation}
The supersymmetric mass term for the chiral multiplets, $m_{ij}$,
transforms as the $\underline{6}$ of $SU(3)$, and the corresponding
supergravity mode appears in the
decomposition of the $\underline{10}\rightarrow
\underline{1}+\underline{6} +\underline{3}$
of $SU(4)$ under $SU(3)\times U(1)$.
In principle, a generic non-zero VEV for $m_{ij}$ will
induce non-zero VEVs for other scalars as well, due to the existence
 of linear couplings of $m$ to other fields in the potential.
However, if we further impose $SO(3)$ symmetry,
by taking an $m_{ij}$ proportional to the identity matrix, a
simple group theory exercise shows that all remaining fields
can be
consistently set to zero\footnote{Unlike other examples \cite{freed1},
the second order perturbation in $m$, which amounts to the trace of
the scalar mass terms, is associated with a stringy state and does not
force us to introduce a second scalar field in the supergravity flow.}.

The five-dimensional action for the scalars,~\cite{grw2},
\beq
L = \sqrt{-g}\left[-{R\over 4}
- {1\over 24}\Tr (U^{-1}\partial U)^2 + V(U)
\right],
\eeq{m32}
is written in terms of a
$27\times 27$ matrix $U$, transforming in the fundamental
representation of
$E_6$ and parametrising the coset $E_6/USp(8)$.
In a unitary gauge, $U$
can be written as $U=e^X, X=\sum_a \lambda_a T_a$,
where $T_a$ are the
generators of $E_6$ that do not belong to $USp(8)$.
This matrix has exactly 42 real independent parameters, which are the
scalars of the supergravity theory.

We shall not discuss here
the philosophy and details of the computation of the potential
and kinetic terms, since the have been extensively described in
several previous papers \cite{dz,grw2,pilch}. In the Appendix, the
interested reader can find a summary of the relevant
formulae and the explicit parametrisation for the coset manifold
representative we used.
The result for the diagonal scalar $m\delta_{ij}$ is\footnote{In the
following we will always set the coupling constant $g$ equal to $2$, so
that the scalar potential in the $N=8$ vacuum, where all scalars have
zero VEV, is normalised as $V(N=8)=-3$.}
\beq
L =\sqrt{-g}\left\{-
{R\over 4} + {1\over 2}(\partial m)^2
- \frac{3}{8} \left[3 +\left(\cosh{{2m\over \sqrt{3}}}\right)^2
+4\cosh{{2m\over \sqrt{3}}}\right]\right\},
\eeq{m33}
where $R$ is the curvature of $AdS_5$.

We want to find a solution of 5d gauged supergravity
with one asymptotically-$AdS_5$ region, corresponding to the UV $N=4$ super
Yang-Mills theory.
The ansatz for the 4-d Poincar{\'e} invariant metric is \cite{gppz2}
\beq
ds^2= dy^2 + e^{2\phi(y)}dx^\mu d x_\mu , \;\;\;\mu =0,1,2,3,
\eeq{m34}
with $\phi \sim y/R + {\rm const}$ when $y\rightarrow \infty$.
We also assume that our scalar only depends on the radial
coordinate. The boundary conditions for the scalars is that they
approach
the $SO(6)$, $N=8$ invariant stationary point: for
$y\rightarrow \infty$, $m$ must vanish.

The solution we are looking for should correspond to an $N=1$
supersymmetric flow on the dual field theory side.
As discussed in \cite{grw2,pilch}, it is possible to determine the
 number
of supersymmetries of a certain supergravity configuration by inspection of
the eigenvalues of the tensor $W_{ab}$ in the scalar potential (see the
Appendix for the definition of the tensors below)
\beq
V = -\frac{1}{8} \left[2W_{ab}W^{ab} - W_{abcd}W^{abcd}\right].
\eeq{m35}
The number of supersymmetries is given by the number of eigenvalues for
which eq.(\ref{I9}) is valid~\cite{freed1}.

In our case, where only the scalar $m$ is considered, $W_{ab}$ has
two different eigenvalues
\bea
W_1&=& -\frac{1}{4} \left(5\cosh{\frac{2m}{\sqrt{3}}}+1 \right),\\
W_2&=& -\frac{3}{4} \left(\cosh{\frac{2m}{\sqrt{3}}} + 1\right),
\eea{36}
with degeneracy $6$ and $2$ respectively. It is easy to check that only
the second one satisfies eq.~(\ref{I9}).
This corresponds to an $N=1$ supersymmetric gauge theory.

Supersymmetry also makes it possible to reduce the problem of solving
the second order equations of motion to solving equations for the fermion
shifts, i.e. first order equations, with $W=-\frac{3}{4}
\left(\cosh{\frac{2m}{\sqrt{3}}} + 1\right)$. They read
\bea
\dot{m}&=& - \frac{\sqrt{3}}{2}\sinh{\frac{2m}{\sqrt{3}}},\\
\dot{\phi}&=& \frac{1}{2}\left(1+ \cosh{\frac{2m}{\sqrt{3}}}\right).
\eea{m37}

These equations can be solved exactly, and give
\bea
\phi(y)&=&\frac{1}{2}\left(y+\log[2\sinh(y-C_1)]\right),\\
m(y)&=&\frac{\sqrt{3}}{2}\log\left[\frac{1+e^{-(y-C_1)}}
{1-e^{-(y-C_1)}}\right].
\eea{m38}

The metric has a singularity at $y=C_1$
\beq
ds^2= dy^2 + |y-C_1| dx^\mu dx_\mu .
\eeq{m39}
Around this point $m$ behaves as
\beq
m \sim -\frac{\sqrt{3}}{2} \log(y-C_1) + \mbox{const}.
\eeq{m310}

Notice that, although singular, this is not the universal behaviour
found in \cite{gppz2}: indeed, because of supersymmetry, it is not
possible to ignore the potential in the equations of motion.
On the other hand, it is easy to see that this solution corresponds
to a true deformation of the gauge theory. Indeed, $m$ approaches the
boundary in the UV ($y \rightarrow \infty$) as $m \sim e^{-y}$, which
is the required behaviour of a deformation (see eq.(\ref{I3})).

As we said before, we expect the gauge theory to exhibit confinement in
the IR. In
order to prove confinement for our solution, we need to check that the Wilson
loop has an area law behaviour. As discussed in the previous section,
to compute the Wilson loop we need to minimise the action for a string
whose endpoints are constrained on a contour $C$ on the boundary.
For the background (\ref{m34}) and with the choice of variables of
eq.(\ref{I15})\footnote{We
consider a rectangular loop $C$ as in eq.~(\ref{I12}) and
we also factorize the trivial dependence in $t$.}, this reads
\beq
E = S/T = \int dx \sqrt{(\partial_x u)^2+f(u)}.
\eeq{m312}

As already explained in the previous Section (see also \cite{gppz2} for
a more detailed discussion), confinement depends on the behaviour
for small $u$ of the function
\beq
f(u) = T^2(u)e^{4\phi(u)}.
\eeq{}
In our example the tension of the fundamental strings and of
the D1-strings, $T(u)$, are, respectively,\footnote{The explicit form of the
matrix $A$
from which the tension can be read is given in the Appendix.}
\bea
T^2_{F1}&=&4\left(\cosh{\frac{4m}{\sqrt{3}}}+\cosh{\frac{2m}{\sqrt{3}}}\right),\\
T^2_{D1}&=&8 \left(\cosh{\frac{m}{\sqrt{3}}}\right)^2,
\eea{m313}
so that the asymptotic behaviour of the corresponding functions $f(u)$ is
\beq
f_{(q\bar{q})}(u) \sim 1, \,\,\,\ f_{(m\bar{m})}(u) \sim \left|u-C_1 \right|.
\eeq{m314}
According to the discussion in the previous section, this is indeed what
we need
to have quark confinement (i.e. a linear potential) and monopole screening.

\subsection{Gaugino Condensate}
Another feature of $N=1$ Super Yang Mills is the presence of a gaugino
condensate. It is then natural to ask whether this feature can be found in
the supergravity solution.
The key point is again the double meaning of the
solutions in AdS: true deformations of the gauge theory correspond to
non-normalizable solutions of the supergravity equation of motion, while
different vacua correspond to normalizable solutions.

In the decomposition $\underline{10}\rightarrow \underline{1}+\underline{6}
+\underline{3}$ of $SU(4)$ under $SU(3)\times U(1)$, a scalar $SU(3)$-singlet
appears: $\sigma$. It corresponds to a
bilinear operator in the gaugino fields \footnote{A non-zero VEV for
this field alone was studied in \cite{gppz1,dz}. It leads to a
non-supersymmetric, conformal IR fixed point.}.

It is tempting to interpret a non-zero VEV for this scalar as the
gaugino condensate. To substantiate this interpretation,
we must look for solutions where both the fields $m$ and
$\sigma$ are given a non-zero VEV and check whether they have the right
UV asymptotic behaviour:  $m$ has to be asymptotic to the
non-normalizable solution
$e^{-(4-\Delta )y}$ of eq.(\ref{I3}), and $\sigma$ to the normalizable
one $e^{-\Delta y}$.
It is easy to see that this is indeed the case.

The five-dimensional Lagrangian for the fields $m$ and $\sigma$ now
becomes
\bea
L &=& \sqrt{-g}\left\{- {R\over 4}
+ {1\over 2}(\partial m)^2+{1\over 2}(\partial \sigma)^2 + \right. \nonumber\\
& & \left. - \frac{3}{8} \left[\left(\cosh{{2m\over \sqrt{3}}}\right)^2
+4\cosh{{2m\over \sqrt{3}}}\cosh{2 \sigma}
-\left(\cosh{2 \sigma}\right)^2 +4 \right]\right\}.
\eea{m315}
The tensor $W_{ab}$ in the potential has one eigenvalue,
$W_2=\frac{3}{4} \left(\cosh{\frac{2m}{\sqrt{3}}} + \cosh{2\sigma}\right)$, 
with degeneracy $2$, satisfying eq.(\ref{I9}). Thus, we are still dealing
with solutions describing $N=1$ supersymmetric RG flows
\bea
\phi(y)&=&\frac{1}{2}\log[2\sinh(y-C_1)] + \frac{1}{6}\log[2\sinh(3y-C_2)],
\label{m316'}\\
m(y)&=&\frac{\sqrt{3}}{2}\log\left[\frac{1+e^{-(y-C_1)}}
{1-e^{-(y-C_1)}}\right], \label{m316''}\\
\sigma(y)&=&\frac{1}{2}\log\left[\frac{1+e^{-(3y-C_2)}}
{1-e^{-(3y-C_2)}}\right].
\eea{m316}

The UV is at $y=+ \infty$, and it is immediate to see that $m$ and
$\sigma$ have the desired asymptotic behaviour for $y \rightarrow \infty$:
\beq
m \sim e^{-y}, \,\,\ \sigma \sim e^{-3y}.
\eeq{m317}

The solution we have found corresponds, therefore, to a supersymmetric
mass deformation of $N=4$ SYM that gives rise to a RG flow to a $N=1$ SYM
vacuum with a gaugino condensate\footnote{Note that the IR
contributions to operator condensates found in \cite{gppz2} vanish in
this case.}.

Finally one may ask whether the presence of the condensate affects the
IR behaviour of the solution, in particular confinement.
The form of the tensions for the F1 and D1-string in the presence
of both $m$ and $\sigma$ is
\bea
T^2_{F1}&=& 4\left[\cosh{\frac{4m}{\sqrt{3}}} +
\cosh{\left(\frac{2m}{\sqrt{3}} +2\sigma \right)} \right],\\
T^2_{D1}&=& 8 \left[\cosh{\left(\frac{m}{\sqrt{3}} - \sigma \right)}\right]^2.
\eea{m318}
It is easy to see from these formulae
that the asymptotic behaviours of the functions
$f_{(q,\bar{q})}(u)$ and $f_{(m,\bar{m})}(u)$ in eq.(\ref{m314}) remain
unchanged: the solution is still confining.

Notice that the only assumption we make is that the integrations constants
$C_{1,2}$ satisfy the inequality $C_2 \leq 3C_1$.

It is interesting to notice that our solution depends on two constants of
integration. This may signal that our supergravity background describes a
two-parameter deformation of pure $N=1$ super Yang-Mills. After all,
the construction in ref.~\cite{w2} shows that such deformation exists,
and is given by a suitable compactification to four dimensions of a
6d $(2,0)$ superconformal theory.

Notice also that eqs.~(\ref{m316'},\ref{m316''},\ref{m316}) describe
chiral-symmetry preserving ($\sigma=0$) as well as chiral-symmetry
breaking flows ($\sigma\neq 0$). The chirality-preserving 
flow is unstable in the
sense that any nonzero $\sigma$ in the UV (i.e. at large positive $y$), no
matter how small, flows in the IR ($y=C_1$) to a finite, nonzero
$\sigma=-\log[\tanh(3C_1/2-C_2/2)]$. This behaviour agrees with field theory
expectations~\cite{ksh,w2}.

Moreover, let us point out that a nonzero condensate $\sigma$ implies the
existence of domain walls, interpolating between vacua with different values
of $\sigma$. Since the tension of these domain walls scales as
N, instead of $\mathrm{N}^2$, in the large-N limit~\cite{ksh},
they do not appear as classical supergravity solitons, but rather as D-branes
\cite{w2}. This is consistent with the vanishing of the potential
barrier between chiral-symmetry breaking vacua, evident from our explicit
solution. To find which quantum soliton or D-brane plays the role of domain
wall in our solution is an interesting problem that we shall leave open
(for the time being, at least).

Finally, we would like to remark that the parametrisation used in
eq.(\ref{m316}) \footnote{See eq.(\ref{aa9}) for the explicit form of
the coset manifold representative.}
allows also to describe
the $N=0$ $SU(3)$ critical point. 
Denoting by $R_{IR}$ the radius of the Anti-de-Sitter space corresponding to
the IR fixed point,
the square mass of the fluctuation of the
field $m$ around the $SU(3)$ point is $-(40/9)R^{-2}_{IR}$, which violates
the unitarity bound \cite{BF}. This supports the
suggestion that the $SU(3)$ critical point is unstable \cite{dzf}. 
Our parametrisation differs from that obtained in
\cite{dz}, which we used in \cite{gppz2}. Thus the scalar
potential in Section 3 of \cite{gppz2} has to be substituted with
eq.(\ref{m33}). This change does not affect the general results discussed in
that paper.

\vskip .2in
\noindent
{\bf Acknowledgements}\vskip .1in
\noindent
L. Girardello, M. Porrati and A. Zaffaroni would
like to thank CERN, where most of this work was done.
M. Porrati would like to thank the ITP at UCSB
for its kind hospitality during completion of this work.
L.G. and
M. Petrini are partially supported by INFN and MURST, and
by the European Commission TMR program ERBFMRX-CT96-0045,
wherein L.G. is associated to the University Torino,
and M. Petrini to the Imperial College, London. M. Porrati
is supported in part by NSF grant no. PHY-9722083.

\section*{Appendix: Conventions and Useful Formulae for the Lagrangian}
\renewcommand{\theequation}{A.\arabic{equation}}
\setcounter{equation}{0}

In this Appendix we discuss some of the details of the computations that
led to the results in the paper. We will present our results for the case
where both the fields $m$ and $\sigma$ are turned on. The case where
only $m$ is considered can be easily recovered by setting $\sigma$ equal to
zero.
This Appendix is not meant to be self-contained, since
the tools for computing the
potential have been already described in details in
\cite{grw2} and recently reviewed in \cite{dz,pilch}.
The  reader may refer to the above-mentioned papers for
more details.

\subsubsection*{Parametrisation of the Coset Manifold}
The five-dimensional Lagrangian for the scalars is written in terms of
the $27 \times 27$ matrix $U$, parametrising the coset $E_6/USp(8)$, as
\beq
L = \sqrt{-g}\left[{R\over 4} - {1\over 24}\Tr (U^{-1}\partial U)^2 + V(U)
\right].
\eeq{aa1}
To evaluate it, one has first to choose a parametrisation for the coset
manifold representative, $U=\exp{X}$, that gives canonical kinetic terms
for the scalars.
The precise form of the matrix $X$ is obtained using the global and
local symmetries of the problem, and the fact that $U$ maps an element of the
representation $\underline{27}$ of $E_6$ into itself.
The general parametrisation for $U$ has been worked out in \cite{grw2}.
In the gauged theory, only the group
$SU(4)\times SL(2;R)$ is a symmetry of the Lagrangian. The $42$ scalars
then decompose according to
\beq
42 \rightarrow 20'_{(0)} + 10_{(-2)} + \bar{10}_{(2)} + 1_{(4)} + 1_{(-4)},
\eeq{aa2}
while the vectors in the $\underline{27}$ decompose as
\beq
27 \rightarrow 15_{(0)} + 6_{(2)} + 6_{(-2)}.
\eeq{aa3}
The subscripts denote the charges of the $U(1)$ factor in $SL(2;R)$.
The $\underline{27}$ of $E_6$ is represented by a couple of antisymmetric
symplectic-traceless indices $A,B$,
running form $1$ to $8$, and, in the $SU(4)\times SL(2;R)$ basis, it
decomposes as
\bea
27 & \rightarrow & 15 + (6,2),\\
z^{AB} & \rightarrow & \left(\begin{array}{c}
                             z_{IJ} \\
                             z^{ I\alpha}
                            \end{array} \right),
\eea{aa4}
with $I,J=1,\dots,6$ indices of $SU(4)$ and $\alpha=1,2$ indices of $SL(2;R)$.
The variation of the vector $z^{AB}$ under $E_{6(6)}$ is:\footnote{The
indices $M,N,..$ are raised with the invariant tensor
$\epsilon_{IJKLMN}=\epsilon^{IJKLMN}$:
$\Sigma^{IJK\alpha}= {1\over{6}}\epsilon^{\alpha\beta}
\epsilon^{IJKLMN}\Sigma_{LMN\beta}$ \cite{grw2}.}
\bea
\delta z_{IJ} &=& -\Lambda^M_I z_{MJ} - \Lambda^M_J z_{IM} +
               \Sigma_{IJM\beta} z^{M\beta},\\
\delta z^{I\alpha} &=& \Lambda^I_M z^{M\alpha} +
\Lambda^{\alpha}_{\beta} z^{I\beta} +
               \Sigma^{MNI\alpha} z_{MN} .
\eea{aa5}
{}From these equations it
is possible to find the form of the matrix $X$ in the $SU(4)\times 
SL(2;R)$ basis:
\beq
X=\left( \begin{array}{cccc}
         -2 \Lambda^{[M}_{[I}\delta^{N]}_{J]}& \Sigma_{IJM\beta} \\
          2\Sigma^{MNI\alpha}& \Lambda^I_J \delta^{\alpha}_{\beta} +
\Lambda^{\alpha}_{\beta} \delta^{I}_{J}
          \end{array} \right).
\eeq{aa6}
Here $\Lambda^{M}_{N}$ and $\Lambda^{\alpha}_{\beta}$ are real and
traceless matrices, that are the generators
of $SU(4)$ and  $SL(2;R)$, while $\Sigma^{MNI\alpha}$ is real and totally
antisymmetric in the indices $MNI$. The physical states are given by the
non-compact generators of $E_{6(6)}$, which correspond to the symmetric
parts of $\Lambda^{M}_{N}$ and $\Lambda^{\alpha}_{\beta}$, and the
self-dual part of  $\Sigma^{MNI\alpha}$. The self-duality condition for
$\Sigma^{MNI\alpha}$ is \cite{grw2}
\beq
\Sigma^\pm _{IJK\alpha}= \pm {1\over{6}}\epsilon_{\alpha\beta}
\epsilon_{IJKLMN}\Sigma^\pm _{LMN\beta}.
\eeq{aa6bis}

In a unitary $USp(8)$ gauge, $U$ can be written as
\beq
U=\exp{X},
\eeq{aa7}
where $X= \sum_{a} \lambda_a T_a$ is given only by the $42$ non compact
generators of $E_6$: these $42$ independent parameters
correspond to the $42$ scalars.

We are interested in the scalars $m$ and $\sigma$ which are
in the $\underline{6}_{(2,-2)}$ of $SU(4)$ and transform as a singlet
and a  $\underline{6}$ of the $SU(3) \subset SU(4)$.
Turning them on breaks $SU(4)$ to $SU(3) \times U(1)$.
We choose the following decomposition of the $\underline{27}$ in  $SU(3)
\times U(1)$ basis \cite{dz}:
\beq
27 \rightarrow \left(1_{(0,0)}, 3_{(4,0)}, \bar{3}_{(-4,0)}, 8_{(0,0)},
3_{(-2,2)},
\bar{3}_{(2,2)}, 3_{(-2,-2)}, \bar{3}_{(2,-2)}\right),
\eeq{aa8}
where the first
index gives the charges under the $U(1)\subset SU(4)$ and the second the
charges under the other $U(1)\subset SL(2;R)$.
With this choice of basis, the matrix $X$ has the following form
\footnote{Notice that the parametrisation we obtained differs slightly
from that given in~\cite{dz}}
\beq
X=\left( \begin{array}{cccccccc}
     0&0&0&0&0&0&0&0 \\
     0&0&0&0&0&{m_{ij}\over{\sqrt{3}}}&\sigma&0 \\
     0&0&0&0&0&\sigma&{\bar{m}_{\bar{\imath}\bar{\jmath}}\over{\sqrt{3}}}&0 \\
     0&0&0&0&\epsilon_{\bar{\jmath}\bar{k}\bar{l}}{m_{il}\over{\sqrt{3}}}&0&0&
        \epsilon_{jkl}{\bar{m}_{\bar{\imath}\bar{l}}\over{\sqrt{3}}} \\
     0&0&0&\epsilon_{jkl}{\bar{m}_{\bar{\imath}\bar{l}}\over{\sqrt{3}}}&0&0&0&0
\\
     0&{\bar{m}_{\bar{\imath}\bar{\jmath}}\over{\sqrt{3}}}&\sigma&0&0&0&0&0 \\
     0&\sigma&{m_{ij}\over{\sqrt{3}}}&0&0&0&0&0 \\
     0&0&0&\epsilon_{\bar{\jmath}\bar{k}\bar{l}}{m_{il}\over{\sqrt{3}}}&0&0&0&0
          \end{array} \right),
\eeq{aa9}\\

where $i,\bar{\imath}=1,\ldots,3$ are complex $SU(3)$ indices.\\

\subsubsection*{Vielbein and Gamma Matrices}
To compute the potential and the tensions of the fundamental and
D1-string we need to know the vielbein $V_{AB}\null^{ab}$,
where $a,b$ are a couple of antisymmetric symplectic-traceless indices
$a,b=1,...,8$, representing
the $\underline{27}$ of $USp(8)$.
This field, being an element of $E_6/USp(8)$, carries both the indices
$A,B=1,\dots,8$ of the $\underline{27}$ of $E_6$ and the
indices $a,b=1,\dots,8$ of the $\underline{27}$ of $USp(8)$.
According to ref.\cite{grw2}, one can go
from the $E_6$ basis to the $USp(8)$ using the $SO(7)$
gamma matrices, $\Gamma$.

In the complex basis we are using, the vielbein is obtained by
multiplying on the right the matrix
$U$  by the following vector of gamma matrices:

\beq
\left(\frac{\gamma_{i\bar{\jmath}}}{2\sqrt{2}},
\frac{\epsilon_{ijk}\gamma_{\bar{\jmath}\bar{k}}}{4\sqrt{2}},
\frac{\epsilon_{\bar{\imath}\bar{\jmath}\bar{k}}\gamma_{jk}}{4\sqrt{2}},
\frac{\gamma_{i\bar{\jmath}}}{2\sqrt{2}},
\frac{\gamma_{i}(1-\Gamma_0)}{4},
\frac{\gamma_{\bar{\imath}}(1-\Gamma_0)}{4},
\frac{\gamma_{i}(1+\Gamma_0)}{4},
\frac{\gamma_{\bar{\imath}}(1+\Gamma_0)}{4}\right).
\eeq{aa10}

The complex gamma matrices, $\gamma_{j}$ and $\gamma_{\bar{\jmath}}$,
have a simple expression in terms of the real matrices $\Gamma$ (here we
used the definition of the $SO(7)$ gamma matrices of ref.~\cite{freed1})
\bea
\gamma_{j}&=&\frac{\Gamma_{j+3} +i\Gamma_j}{\sqrt{2}},\\
\gamma_{\bar{\jmath}}&=&\frac{\Gamma_{j+3} -i\Gamma_j}{\sqrt{2}},
\eea{aa11}
where $j,\bar{\jmath}=1,2,3$.
\subsubsection*{Potential}
The scalar potential is expressed in terms of the two $USp(8)$ tensors
\bea
W_{abcd}&=&\epsilon^{\alpha \beta}\eta^{IJ}V_{I\alpha ab}V_{J\beta ab},\\
W_{ac}&=&-i \left(\Gamma_0\right)^{db}W_{abcd},
\eea{aa12}
as \cite{grw2}
\beq
V = -\frac{1}{8} \left[2W_{ab}W^{ab} - W_{abcd}W^{abcd}\right],
\eeq{aa13}
where the $USp(8)$ indices are raised and lowered with the
matrix $\Gamma_0$, as shown in ref.~\cite{grw2}.
Once the vielbein are known, its
evaluation is a lengthy but straightforward computation. Plugging
$V_{AB}\null^{ab}$ in the expressions (\ref{aa10}),(\ref{aa11}) and
performing some gamma-matrix algebra one gets:
\beq
V = - \frac{3}{8} \left[\left(\cosh{{2m\over \sqrt{3}}}\right)^2
+4\cosh{{2m\over \sqrt{3}}}\cosh{2 \sigma}
-\left(\cosh{2 \sigma}\right)^2 +4 \right].
\eeq{aa14}
As already mentioned in the paper, the eigenvalues of the tensor
$W_{ab}$ are related to the number of supersymmetries of the RG flows.
In our case $W_{ab}$ reads
\beq
W_{ab}= -\frac{1}{16}\left[\left(18 I +
J\right)\cosh{\frac{2m}{\sqrt{3}}}
+ \left(6 I - J\right)\cosh{2\sigma}\right],
\eeq{aa15}
and it has two eigenvalues
\bea
W_1 &=&-\frac{1}{4}\left(5\cosh{\frac{2m}{\sqrt{3}}}
+ \cosh{2\sigma}\right),\\
W_2&=&-\frac{3}{4}\left(\cosh{\frac{2m}{\sqrt{3}}} +\cosh{2\sigma}\right),
\eea{aa16}
with degeneracy $6$ and $2$ respectively. In eq.(\ref{aa15}), $I$ is the
$(8 \times 8)$ identity matrix and $J=2{\rm \, diag}(1,1,-3,1,1,1,-3,1)$.

\subsubsection*{Kinetic Term for the $B$ Fields}
Finally, we give the explicit form of the matrix $A_{I\alpha,J\beta}$ in
eq.(\ref{I13}). The kinetic terms for the $12$ antisymmetric tensors
$B_{\mu \nu}^{I\alpha}$ was given in \cite{grw2}
\beq
-\frac{1}{8}B_{\mu \nu ab}B^{\mu \nu ab} +
\frac{1}{8}\epsilon^{\mu \nu\rho\sigma\tau}
\eta_{IJ}\epsilon_{\alpha\beta}
B_{\mu \nu}^{I\alpha}D_{\rho}B_{\sigma\tau}^{J\beta},
\eeq{aa17}
with $B_{\mu \nu}^{ab}=B_{\mu \nu}^{I\alpha}V_{I\alpha}^{ab}$.
The form in eq.(\ref{I13}) is obtained when we explicit the dependence
on the vielbein in the first term, and we define the matrix
\beq
 A_{I\alpha,J\beta}=V_{I\alpha}\null^{ab}V_{J\beta ab}.
\eeq{aa18}
In our example it has the form:
\beq
\left( \begin{array}{cccc}
  \cosh^2{\left({m\over{\sqrt{3}}}-\sigma\right)}
   &0&0&
  -\sinh^2{\left({m\over{\sqrt{3}}}-\sigma\right)}\\
   0&\sinh^2{{2m\over{\sqrt{3}}}}+\cosh^2{\left({m\over{\sqrt{3}}}
+\sigma\right)}&
   \cosh^2{\left({m\over{\sqrt{3}}}+\sigma\right)}
-\cosh^2{{2m\over{\sqrt{3}}}}&0\\
   0&\cosh^2{\left({m\over{\sqrt{3}}}+\sigma\right)}
-\cosh^2{{2m\over{\sqrt{3}}}}&
    \sinh^2{{2m\over{\sqrt{3}}}}+
      \cosh^2{\left({m\over{\sqrt{3}}}+\sigma\right)}&0\\
    - \sinh^2{\left({m\over{\sqrt{3}}}-\sigma\right)}
   &0&0&
  \cosh^2{\left({m\over{\sqrt{3}}}-\sigma\right)}
          \end{array} \right). \\
\eeq{aa19}

\end{document}